%% file: main.tex
\documentclass[10pt, paper=a4, UKenglish]{article}
\usepackage{graphicx}
\def\Title#1{\begin{center} {\Large #1 } \end{center}}
\def\Author#1{\begin{center}{ \sc #1} \end{center}}
\def\Address#1{\begin{center}{ \it #1} \end{center}}

\newcommand\pubblock{\rightline{\begin{tabular}{l} Proceedings of the CTD/WIT 2019\\ \pubnumber\\
         \pubdate  \end{tabular}}}

\newenvironment{Abstract}{\begin{quotation} \begin{center} 
             \large ABSTRACT \end{center}\bigskip 
      \begin{center}\begin{large}}{\end{large}\end{center} \end{quotation}}

\newenvironment{Presented}{\begin{quotation} \begin{center} 
             PRESENTED AT\end{center}\bigskip 
      \begin{center}\begin{large}}{\end{large}\end{center} \end{quotation}}

\input econfmacros.tex

\usepackage[utf8]{inputenc}
\usepackage{booktabs}
\usepackage{lineno}
\usepackage{placeins}
\usepackage{subfig}
\usepackage{hyperref}
\usepackage{xcolor}

\newcommand\pubnumber{PROC-CTD19-014}

\newcommand\pubdate{\today}

\def\affiliation{
$^1$CERN, $^2$Luleå University of Technology, $^3$DESY
}

\newcommand{\conference}{Connecting the Dots and Workshop on Intelligent Trackers (CTD/WIT 2019)\\
Instituto de F\'isica Corpuscular (IFIC), Valencia, Spain\\ 
April 2-5, 2019}

\usepackage{fancyhdr}
\definecolor{mygrey}{RGB}{105,105,105}
\begin{document}










\large
\begin{titlepage}
\pubblock

\vfill
\Title{Towards Fast Displaced Vertex Finding}
\vfill

\Author{\underline{Kim Albertsson}$^{1, 2}$ Federico Meloni$^{3}$}
\Address{\affiliation}
\vfill

\begin{Abstract}
Many Standard Model extensions predict metastable massive particles that can be detected by looking for displaced decay vertices in the inner detector volume. Current approaches to search for these events in high-energy particle collisions rely on the presence of additional energetic signatures to make an online selection during data-taking, as the reconstruction of displaced vertices is computationally intensive. Enabling trigger-level reconstruction of displaced vertices could significantly enhance the reach of such searches.


This work is a first step approximating the location of the primary vertex in an idealised detector geometry using a 4-layer dense neural networks for regression of the vertex location yielding a precision of $O(1\ \mathrm{mm})$ [$O(20\ \mathrm{mm})$] RMS in a low [high] track multiplicity environment.

\end{Abstract}

\vfill

\begin{Presented}
\conference
\end{Presented}
\vfill
\end{titlepage}
\def\thefootnote{\fnsymbol{footnote}}
\setcounter{footnote}{0}
%

\normalsize 


\input{meat.tex}


\end{document}

%% file: econfmacros.tex
\def\beq{\begin{equation}}
\def\eeq#1{\label{#1}\end{equation}}
\def\eeqn{\end{equation}}

\def\beqa{\begin{eqnarray}}
\def\eeqa#1{\label{#1}\end{eqnarray}}
\def\eeqan{\end{eqnarray}}

\let\bar=\overbar

\def\Dslash{\not{\hbox{\kern-4pt $D$}}}
\def\dslash{\not{\hbox{\kern-2pt $\del$}}}

\def\msb{{\bar{\ssstyle M \kern -1pt S}}}



%% file: meat.tex
\newcommand{\eeZmm}{$e^+e^- \rightarrow Z \rightarrow \mu^+\mu^-$\ }
\newcommand{\ppZmm}{$pp \rightarrow Z \rightarrow \mu^+\mu^-$\ }
\newcommand{\pptt}{$pp \rightarrow t\bar{t}$\ }

\section{Introduction}
\label{intro}

Despite the experimental and theoretical success of the Standard Model of particle physics, gaps still exist. Many beyond Standard Model (BSM) theories addressing these gaps have been proposed, some predicting new particles that can appear at current colliders. However, the particles remain elusive spurring development of specialised detection techniques.

One area of exploration is the search for long-lived particles (LLPs) \cite{bsm}. There particles have an expected life-time long enough such that their decay occurs with a significant displacement from the location where they are produced. Of particular interest for this work are those particles that decay inside the detector volume, possibly giving rise to \emph{displaced vertices}.

Traditional reconstruction techniques rely on explicit track reconstruction before fitting vertices and are optimised for primary vertex reconstruction and short-lived particles reducing their efficiency for displaced vertices. Such techniques can be adapted to finding displaced decays \cite{atlas-lrt}\cite{cms-lrt}, but the resulting algorithm is unsuitable for online application due to high computational costs.

This work explores alternative techniques based on deep learning to reconstruct vertex locations directly from raw detector hits for application at the online trigger level.



To reduce computational overhead, the problem is divided into three steps: (i) Estimate the position of primary interaction; (ii) Define a region of interest based on trigger-level objects (e.g. high-energy muons); (iii) Find displaced vertices in constrained search space.


This work investigates the first step, primary vertex regression from detector hit data. The investigation is limited to feed-forward linear networks because of their small computational cost and thus suitability for integration in an online software trigger. Feed-forward networks are the simplest neural network approach, a method that can utilise hierarchical representation, enabling it to potentially discover indirect correlations in the data.



\section{Data sets and generation}
\label{sec:data}
A custom event simulation was developed to study the problem using MadGraph5\cite{madgraph} and Pythia\cite{pythia} for event generation and includes both primary and soft interactions (pile-up). Delphes\cite{delphes} was used for particle propagation, with one custom module for simulation in a homogeneous magnetic field and one to implement the detector geometry. For this study the exact intersections between particle tracks and active detector surfaces was used. This implies that exact track parameters can be recovered from the data.

\begin{figure}[!htb]
  \centering
  \subfloat[View in the \mbox{xy-plane}.]{\includegraphics[width=0.49\linewidth]{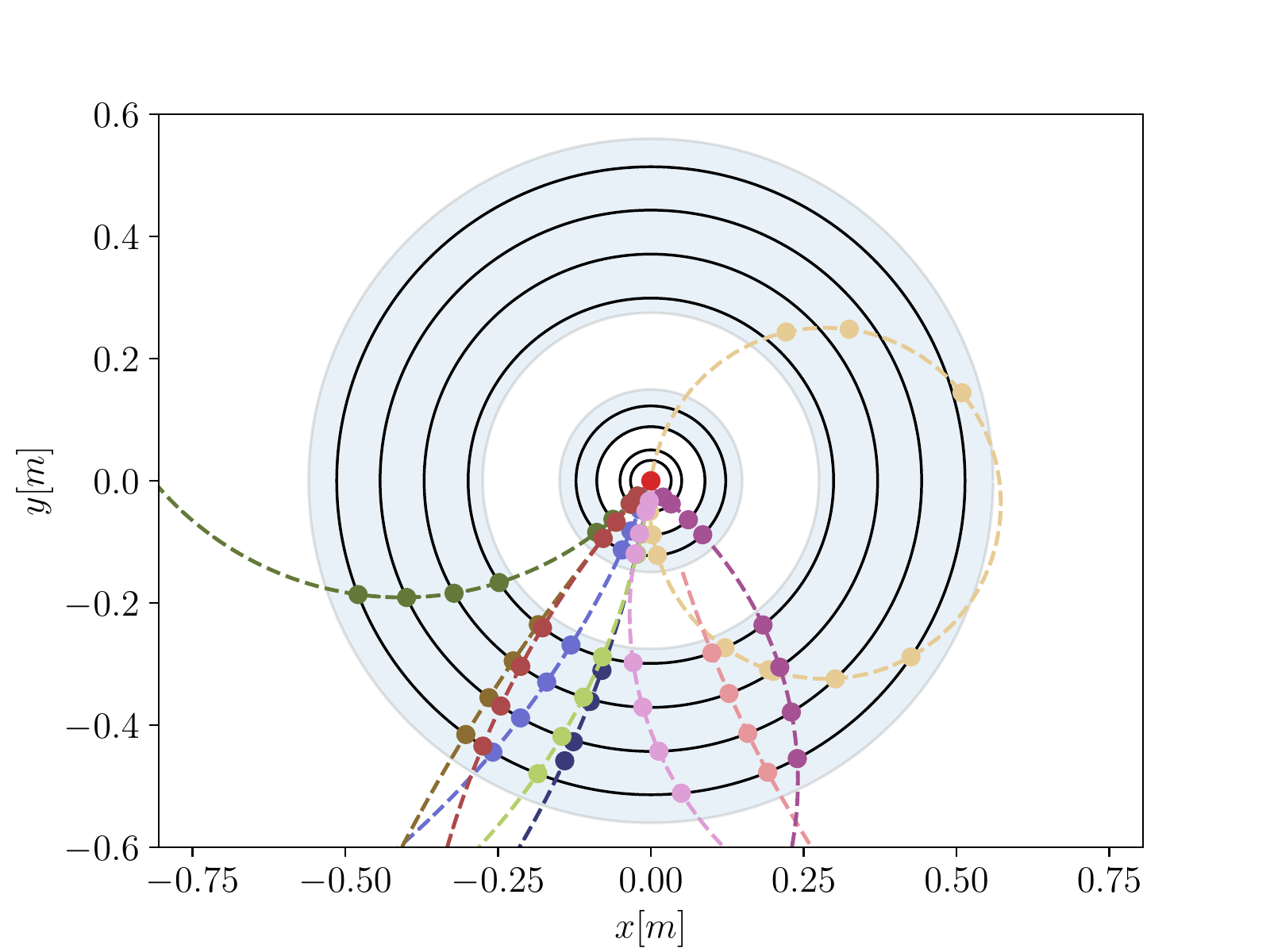}}
  \subfloat[View in the \mbox{zr-plane}.]{\includegraphics[width=0.49\linewidth]{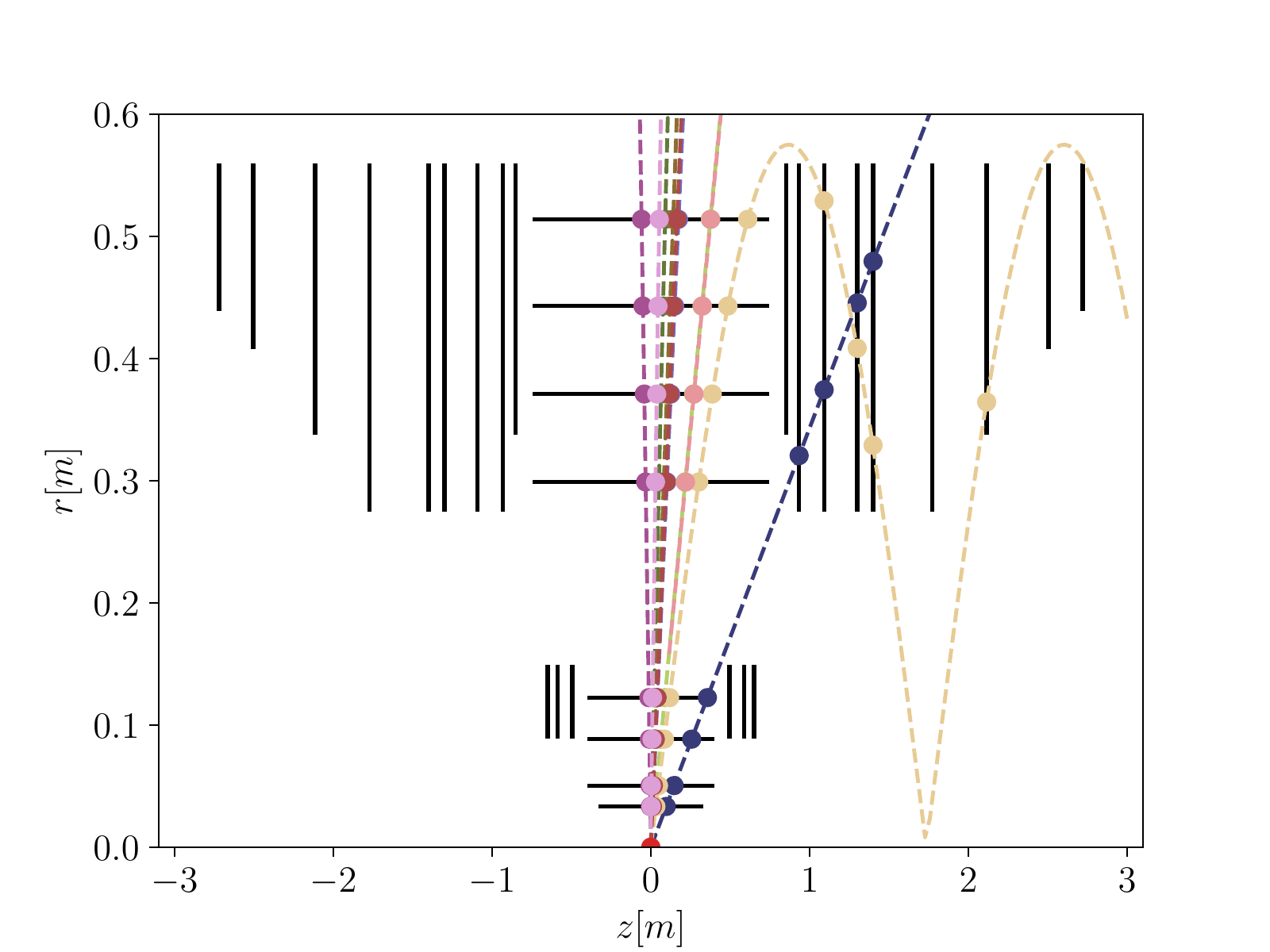}}
  \caption{Example proton-proton event viewed with simulated detector. Dots represent detector hits, dotted lines represent the true track trajectory. Thick black lines correspond to detector surfaces. The light blue shaded area corresponds to endcap regions. The primary vertex is shown as a red dot.}
  \label{fig:event}
\end{figure}

The simulation stores simulated particle track parameters and the three coordinates of each detector hit, $r, \varphi, z$. The detector geometry consists of eight barrel layers and in total 13 endcap layers, shown in \autoref{fig:event}, approximating the pixel and SCT subsystems of the ATLAS inner detector.  The detector surfaces are modelled as ideal shapes.

\paragraph{}
For this investigation, three different data sets were generated containing $200'000$ events each:
\begin{itemize}
\setlength\itemsep{-0.5em}
    \item \eeZmm --- Electron-positron collisions produce clean events with small track multiplicities, used to study efficiency and biases.
    \item \ppZmm, and \pptt --- Two proton-proton processes are studied. These collisions are generally busier with larger track multiplicities. Used to study robustness under pile-up. In the generated data, the mean number of tracks per event was 190 (250) for the Z boson (top pair) process.
\end{itemize}



\section{Experimental setup}

To establish a baseline model the data were preprocessed using the following two steps:
\begin{itemize}
\setlength\itemsep{-0.5em}
    \item For each event, the first $N_t$ truth-level tracks are selected. Tracks are sorted according to their initial direction $\phi_{track}$. If fewer than $N_t$ tracks are available in the event the vector is zero-padded at this stage. $N_t$ was chosen as 2 for the electron data, and 200 for the proton processes, close to the mean number of tracks for a typical event.
    \item For each selected track, a fixed number of hits, $N_h$, were selected and sorted \emph{in order of track propagation}. In experiments, $N_h$ was selected to be large enough to cover a typical track.
\end{itemize}

The number of features per input hit, $N_f$, is 3.


\paragraph{}


The neural network hyper-parameter optimisation was performed using the \eeZmm data. A grid search was used to perform the optimisation with a total of 1296 evaluated points using a regular logarithmic spacing for between 8 and 1024 hidden units per layer. The process and pile-up dependency was investigated using all three data sets. For both setups, the data were split into training data and validation data using half the data for each split. Early stopping was used for regularisation. Pile-up, when included, followed a Poisson distribution with $\mu=1$.

\paragraph{}
The configurations are summarised in \autoref{tab:arch} where $\Pi$ gives the total network input size and is calculated as $\Pi=N_tN_hN_f$.


\begin{table}[ht]
\centering
\begin{tabular}{l cccc cccc c}
    \toprule
        & \multicolumn{4}{c}{\textbf{Input size}}
        & \multicolumn{4}{c}{\textbf{Layer size}}
        & \textbf{Output} \\
    
        \cmidrule(lr){2-5} \cmidrule(lr){6-9}
    
        & $N_t$ & $N_h$ & $N_f$ & $\Pi$
        & 1 & 2 & 3 & 4 &
        \textbf{size}\\
    
    \midrule
    (i) Hyper-param.  & 2   & 14 & 3 & 84   & 32-1024 & 32-1024 & 8-256   & 8-256   & 1 \\
    (ii) Process dep. & 200 & 10 & 3 & 6000 & 512     & 512     & 256     & 256     & 1 \\
    \bottomrule
\end{tabular}
\caption{Network configuration for the two experiment setups.}
\label{tab:arch}
\end{table}



\FloatBarrier
\section{Results}

\begin{figure}
  \centering
  \includegraphics[width=0.80\linewidth]{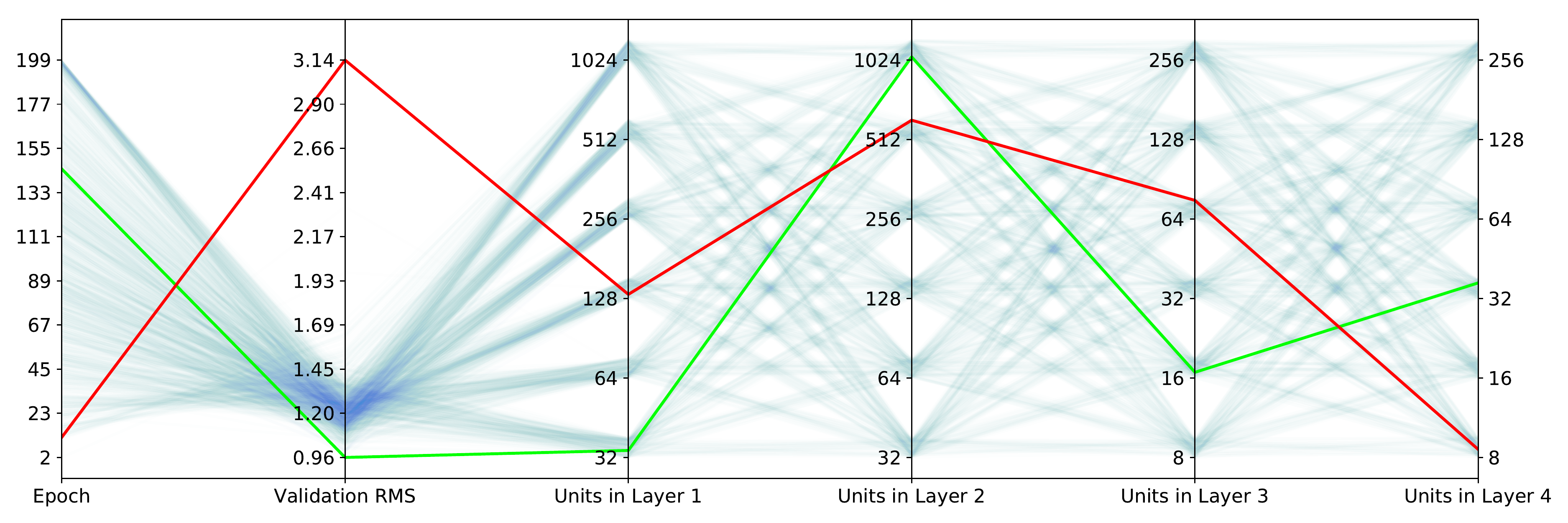}
  \caption{Parallel axis plot of different configurations of hyper-parameters trained on the \eeZmm data set. The first axis shows the number of training epochs. The second axis shows the performance of the configuration as the average reconstruction error across all studied events. The last four axes show the number units in each of the layers.
  The green line represents the best performing model and the red line represents the worst performing model. Blue lines are less transparent with better performance.
  }
  \label{fig:hyperopt}
\end{figure}


The results for the hyper-parameter optimisation on the \eeZmm data is shown in  \autoref{fig:hyperopt}. The best performing point yielded an RMS of 0.92 mm in $z$, the average was 1.2 mm and the worst performing model had an RMS of 3.1 mm.

\begin{table}
\centering
\begin{tabular}{l cc cc cc}
    \toprule
    & \multicolumn{2}{c}{\textbf{\eeZmm}}
    & \multicolumn{2}{c}{\textbf{\ppZmm}}
    & \multicolumn{2}{c}{\textbf{\pptt}} \\
    \cmidrule(lr){2-3}\cmidrule(lr){4-5}\cmidrule(lr){6-7}
    & bias & RMS & bias & RMS  & bias & RMS \\
    \midrule
    No pile-up        & $-0.10$ mm & 0.98 mm & 0.92 mm & 20 mm & 1.5 mm & 16 mm \\
    Pile-up ($\mu=1$) & ---      & ---     & 4.0  mm & 28 mm & 3.1 mm & 22 mm \\
    \bottomrule
\end{tabular}
\caption{Average of the squared error as bias, and the root mean square measure of the error as RMS when regressing the primary vertex $z$ location from raw detector hits.}
\label{tab:result}
\end{table}

Results for the different physics processes considered are shown in \autoref{tab:result}. The reconstruction performs best at low track multiplicities. For larger multiplicities the performance is degraded, with a larger degradation for the Z boson sample as compared to the top pair one.

The time taken for evaluating a single event on a CPU was on the order of 0.10 ms, sufficiently fast for the inclusion in a software trigger, which often operate at time scales of several hundred ms.



The data preprocessing is highly idealised. The network is fed fixed length tracks with hits in track order, furthermore the tracks are sorted. Additionally, the hits use the exact detector surface crossing, meaning track parameters are exactly recoverable. This point coupled with the fact that different assignments of hyper parameters perform similarly suggests a lack of model capacity.

\FloatBarrier
\section{Conclusion}
An initial study of the regression of the $z$-coordinate of the primary vertex from detector hits was investigated using a 4-layer feed-forward linear network. The results show that using this setup a precision of $O(1\ \mathrm{mm})$ RMS can be reached in an idealised low track multiplicity setting. The performance is degraded to $O(20\ \mathrm{mm})$ RMS for processes with a track multiplicity on the order of 200.

The results highlight a limited modelling capacity in the currently considered network architecture and future work will focus on finding a model better suited for the problem setup.


